\documentclass[12pt]{article}
 \usepackage{graphicx,amsmath,amssymb}
\usepackage{indentfirst}
 \usepackage[usenames]{color}
 \usepackage[colorlinks=true, urlcolor=navyblue, linkcolor=navyblue, citecolor=navyblue]{hyperref}
\usepackage{epstopdf}
\usepackage{appendix}

\definecolor{navyblue}{rgb}{0,0.08,0.45}

\def\Dslash{\raise.15ex\hbox{/}\kern-.7em D}
\def\Pslash{\raise.15ex\hbox{/}\kern-.7em P}

\newcommand{\beq}{\begin{equation}}
\newcommand{\enq}{\end{equation}}
\newcommand{\beqa}{\begin{eqnarray}}
\newcommand{\beqast}{\begin{eqnarray*}}
\newcommand{\enqa}{\end{eqnarray}}
\newcommand{\enqast}{\end{eqnarray*}}
\newcommand{\beml}{\begin{multline}}
\newcommand{\enml}{\end{multline}}
\newcommand{\nn}{\nonumber}
\newcommand{\req}[1]{(\ref{#1})}

\newcommand{\pa}{\partial}

\newcommand{\bec}{\begin{center}}
\newcommand{\enc}{\end{center}}
\newcommand{\beqo}{\begin{quote}}
\newcommand{\enqo}{\end{quote}}

\newcommand{\half}{{\textstyle{\frac{1}{2}}}}
\newcommand{\quart}{{\textstyle{\frac{1}{4}}}}

\newcommand{\ze}{\zeta}

\newcommand{\la}{\lambda}

\newcommand{\vp}{\varphi}

\begin{document}

\begin{flushright}
{
\small
SLAC--PUB--15366\\
\date{today}}
\end{flushright}

\vspace{60pt}

\centerline{\Large \bf Threefold Complementary Approach}

\vspace{10pt}

\centerline{\Large \bf to Holographic QCD}

\vspace{20pt}

\centerline{{
Stanley J. Brodsky,$^{a}$ 
\footnote{E-mail: \href{mailto:sjbth@slac.stanford.edu}{sjbth@slac.stanford.edu}}
Guy F. de T\'eramond,$^{b}$ 
\footnote{E-mail: \href{mailto:gdt@asterix.crnet.cr}{gdt@asterix.crnet.cr}}
and
Hans G\"unter Dosch$^{c}$ 
\footnote{E-mail: \href{mailto:gdt@asterix.crnet.cr}{dosch@thphys.uni-heidelberg.de}}
}}

\vspace{30pt}

{\centerline {$^{a}${\it SLAC National Accelerator Laboratory, 
Stanford University, Stanford, CA 94309, USA}}

\vspace{4pt}

{\centerline {$^{b}${\it Universidad de Costa Rica, San Jos\'e, Costa Rica}}

\vspace{4pt}

{\centerline {$^{c}${\it Institut f\"ur Theoretische Physik, Philosophenweg 16, D-6900 Heidelberg, Germany}}

 \vspace{60pt}

\begin{abstract}

A complementary  approach,  derived from (a) higher-dimensional anti--de Sitter (AdS) space, (b) light-front quantization and  (c) the invariance properties of the full conformal group in  one dimension  leads to a nonperturbative relativistic light-front wave equation which incorporates 
essential spectroscopic and dynamical features of hadron physics.  The fundamental  conformal symmetry of the classical QCD Lagrangian in the limit of massless quarks is encoded in the resulting effective theory. The mass scale for confinement emerges from the isomorphism between the conformal group and $SO(2,1)$.
This scale appears in the light-front Hamiltonian by mapping to the evolution operator in the formalism  of de Alfaro, Fubini and Furlan,  which retains the conformal invariance of the action.  Remarkably, the specific form of the  confinement interaction and the corresponding  modification of AdS space are uniquely determined in this procedure.

\end{abstract}

\newpage

\section{Introduction}

One of the most interesting questions in hadron physics is the nature of the interaction which permanently confines quarks and gluons, the fundamental fields of QCD.  In the chiral limit no mass scale appears in the classical QCD Lagrangian, and its action is conformally invariant.   Thus a second fundamental  question is what sets the value of the mass scale of confining dynamics corresponding to the  parameter $\Lambda_{\rm QCD} $ that appears in quantum loops.

The AdS/CFT correspondence between gravity on a higher-dimensional anti--de Sitter (AdS) space and conformal field theories (CFT) in physical space-time~\cite{Maldacena:1997re}
is an explicit realization of the holographic principle, which postulates that a gravitational system may  be equivalent to a non-gravitational system in one fewer dimension  \cite{'tHooft:1993gx, Susskind:1994vu}. The correspondence has led to a semiclassical approximation for strongly-coupled quantum field theories which provides physical insights into its nonperturbative dynamics.  In practice, the duality provides an effective gravity description in a ($d+1$)-dimensional AdS space-time in terms of a flat $d$-dimensional conformally-invariant quantum field theory defined on the AdS asymptotic boundary~\cite{Gubser:1998bc, Witten:1998qj}.  Thus, in principle, one can compute physical observables in a strongly coupled gauge theory  in terms of a weakly coupled classical gravity theory.

Recent analytical insights into the nonperturbative nature of the confining interaction follow from the remarkable holographic correspondence between the equations of motion in  AdS space  and the light-front (LF) Hamiltonian equations of motion for relativistic light hadron bound-states in physical space-time~\cite{deTeramond:2008ht}.   
In fact, the mapping of the equations of motion~\cite{deTeramond:2008ht}  and the matching of the  electromagnetic~\cite{Brodsky:2006uqa, Brodsky:2007hb} and gravitational~\cite{Brodsky:2008pf} form factors in AdS space~\cite{Polchinski:2002jw, Abidin:2008ku} with the corresponding expressions  derived from light-front quantization in physical space time is the central feature of the light-front holographic approach to hadronic physics.  This  approach allows us to establish a precise relation between wave functions in AdS space and the light-front wavefunctions describing the internal structure of hadrons.

To a first semiclassical approximation, light-front  QCD  is formally equivalent to the equations of motion on a fixed gravitational background asymptotic to AdS$_5$.   In this semiclassical approach to AdS gravity, the confinement properties can  be encoded in a dilaton  profile $\vp(z)$ in AdS space which introduces a length scale in the AdS action and leads to conformal symmetry breaking, but its form is left largely unspecified. The introduction of a dilaton profile is equivalent to a modification of the AdS metric, even for arbitrary spin~\cite{deTeramond:2013it}.

In this Letter we shall  show that the form of the confining interaction appearing in the light-front  bound-state wave equations  for hadrons  follows from  an  elegant 
construction of Hamilton operators  pioneered by V.~de~Alfaro, S.~Fubini and G.~Furlan  (dAFF)~\cite{deAlfaro:1976je},  based on the  invariance properties of a field theory in one dimension under the general group of conformal  transformations.    The construction of  dAFF retains conformal invariance of the action despite the introduction of a fundamental length scale  in the light-front Hamiltonian.  Remarkably, the specific form of the  confinement interaction  in the light-front equation of motion is uniquely determined by this procedure.
In fact, by combining  analyses  from  AdS space,  LF quantization,  and the  dAFF construction we will show that the form of the dilaton in AdS space is uniquely determined to have the form $\varphi(z) = \lambda \, z^2$ which leads to linear Regge trajectories~\cite{Karch:2006pv} and a massless pion~\cite{deTeramond:2009xk}.

The mass scale $\sqrt{\lambda}$ for hadrons appears from breaking of dilation invariance in the light-front Hamiltonian,  preserving the conformal invariance of the action.  The result is a relativistic and frame-independent semiclassical wave equation, which predicts a massless pion in the massless quark  limit and Regge behavior for hadrons with the same slope in both orbital angular momentum $L$ and the radial quantum number $n$~\cite{deTeramond:2009xk}.  The  model  also predicts hadronic light-front wavefunctions which underlie form factors~\cite{deTeramond:2012rt} and other dynamical observables, as well as vector meson electroproduction~\cite{Forshaw:2012im}.

\section{Light-Front Dynamics}

The central problem for solving QCD using the LF Hamiltonian method can be reduced to the derivation of the effective interaction in the LF wave equation which acts on the valence sector of the theory and has, by definition, the same eigenvalue spectrum as the initial Hamiltonian problem.  In order to carry out this program one must systematically express the higher Fock components as functionals of the lower ones. The method has the advantage that the Fock space is not truncated and the symmetries of the Lagrangian are preserved~\cite{Pauli:1998tf}.

In the limit of zero quark masses the longitudinal modes decouple  from the  invariant  LF Hamiltonian  equation  $H_{LF} \vert \phi \rangle  =  M^2 \vert \phi \rangle$
with  $H_{LF} = P_\mu P^\mu  =  P^- P^+ -  \vec{P}_\perp^2$. The generators $P = (P^-, P^+,  \vec{P}_\perp)$, $P^\pm = P^0 \pm P^3$,  are constructed canonically from the QCD Lagrangian by quantizing the system on the light-front at fixed LF time $x^+$, $x^\pm = x^0 \pm x^3$~\cite{Brodsky:1997de}. The LF Hamiltonian $P^-$ generates the LF time evolution 
\beq \label{Pm}
P^- \vert \phi \rangle = i \frac{\pa}{\pa x^+} \vert \phi \rangle,
\enq
whereas the LF longitudinal $P^+$ and transverse momentum $\vec P_\perp$ are kinematical generators.
We obtain the wave equation~\cite{deTeramond:2008ht}
\beq \label{LFWE}
\left(-\frac{d^2}{d\ze^2}
- \frac{1 - 4L^2}{4\ze^2} + U\left(\ze, J\right) \right)
\phi_{n,J,L} = M^2 \phi_{n,J,L},
\enq
a relativistic single-variable  LF Schr\"odinger equation.  
This equation describes the spectrum of mesons as a function of $n$, the number of nodes in $\zeta$, the total angular momentum  $J$, which represents the maximum value of $\vert J^3 \vert$, $J = \max \vert J^3 \vert$,
and the internal orbital angular momentum of the constituents $L= \max \vert L^3\vert$.
The variable $z$ of AdS space is identified with the LF   boost-invariant transverse-impact variable $\zeta$~\cite{Brodsky:2006uqa}, 
thus giving the holographic variable a precise definition in LF QCD~\cite{deTeramond:2008ht, Brodsky:2006uqa}.
 For a two-parton bound state $\zeta^2 = x(1-x) b^{\,2}_\perp$,
where $x$ is the longitudinal momentum fraction and $ b_\perp$ is  the transverse-impact distance between the quark and antiquark. 
In the exact QCD theory the interaction $U$ in \req{LFWE} is related to the two-particle irreducible $q \bar q$ Green's function.  It represents the equal LF time $q \bar q$ potential which appears after reducing out the infinite tower of higher Fock states.

Recently we have derived wave equations for hadrons with arbitrary spin starting from an effective action in  AdS space~\cite{deTeramond:2013it}.    An essential element is the mapping of the higher-dimensional equations  to the LF Hamiltonian equation  found in Ref.~\cite {deTeramond:2008ht}.  This procedure allows a clear distinction between the kinematical and dynamical aspects of the LF holographic approach to hadron physics.  Accordingly, the non-trivial geometry of pure AdS space encodes the kinematics,  and the additional deformations of AdS encode the dynamics, including confinement~\cite{deTeramond:2013it}.

The existence of a weakly coupled classical gravity with negligible  quantum corrections requires that the corresponding  dual field theory has a large number of degrees of freedom. 
In the  prototypical  AdS/CFT duality~\cite{Maldacena:1997re} this is realized by the large $N_C$ limit. In the light-front,  this limit is not a natural concept, and the LF mapping is carried out for $N_C = 3$, with remarkable phenomenological success. Following the original holographic ideas~\cite{Bekenstein:1973ur, Hawking:1974sw} it is tempting to conjecture that the required large number of degrees of freedom is provided by the  strongly correlated multiple-particle states in the Fock expansion in light-front dynamics. In fact,
in the light-front approach, the effective potential  in the LF Schr\"odinger equation \req{LFWE} is the result of integrating out all higher Fock states, corresponding to an infinite number of degrees of freedom. This is apparent, for example, if one identifies the sum of infra-red sensitive ``H'' diagrams as the source of the effective potential, since the horizontal rungs correspond to an infinite number of higher gluonic Fock states~\cite{Appelquist:1977es}. The reduction of higher Fock states to an effective potential is not related to the value of $N_C$~\footnote{An interesting connection of the AdS/QCD duality using Ehrenfest's correspondence principle between classical and quantum mechanics in the context of light-front relativistic dynamics is given in Ref.~\cite{Glazek:2013jba}.}.

For $d=4$ one finds  from the dilaton-modified AdS action the LF potential~\cite{deTeramond:2010ge,deTeramond:2013it}
\beq \label{U}
U(\ze, J) = \frac{1}{2}\vp''(\ze) +\frac{1}{4} \vp'(\ze)^2  + \frac{2J - 3}{2 \zeta} \vp'(\ze) ,
\enq
provided that the product of the AdS mass $m$ and the  AdS curvature radius $R$ are related to the total and orbital angular momentum, $J$ and  $L$ respectively, according to $(m  R)^2 = - (2-J)^2 + L^2$.  The critical value  $J=L=0$  corresponds to the lowest possible stable solution, the ground state of the LF Hamiltonian, in agreement with the AdS stability bound   $(m R)^2 \ge - 4$~\cite{Breitenlohner:1982jf}.

The choice $\vp(z) = \la z^2$, as demanded by the construction described below,  leads through \req{U} to the  LF potential
\beq \label{UU}
U(\ze, J) =   \la^2 \ze^2 + 2 \la (J - 1),
\enq
and  Eq.~(\ref{LFWE}) yields  for $\lambda >0$
 a mass spectrum for mesons characterized by the total angular momentum $J$, the orbital angular momentum $L$ and orbital excitation $n$
\beq
M_{n, J, L}^2 = 4 \la \left(n + \frac{J+L}{2} \right).
\enq
This result not only implies linear Regge trajectories, but also a massless pion and the relation between the $\rho$ and $a_1$ mass usually obtained from the 
Weinberg sum rules~\cite{Weinberg:1967kj}
\beq
m_\pi^2 = M_{0,0,0}^2 = 0, \,
m_\rho^2 = M_{0,1,0}^2 = 2 \la, \,
m_{a_1}^2 = M_{0,1,1}^2 = 4 \la.
\enq

\section{Conformal Quantum Mechanics}

 In the following we shall show that the uniqueness of the dilaton profile and the quadratic LF potential $ \la^2 \ze^2$
 follows  from the underlying conformality of the theory according to the algebraic construction of dAFF~\cite{deAlfaro:1976je}.
 In fact, the LF Hamiltonian can also be derived 
from a one-dimensional quantum field theory (conformal quantum mechanics)  starting from the action~\cite{deAlfaro:1976je} 
\beq \label{S}
S = \half \int dt \Big( \dot Q^2  - \frac{g}{Q^2} \Big),
\enq
where  $\dot Q = dQ/dt$ and $g$ is a dimensionless number 
 (the dimension of $Q$ is $[Q] = [t^{1/2}]$). The equation of motion 
\beq \label{eom}
\ddot{Q} - \frac{g}{Q^3} = 0,
\enq
and the generator of evolution in the variable $t$, the Hamiltonian 
\beq \label{Htq}
H_t = \half \Big(\dot Q^2 + \frac{g}{Q ^2}\Big),
\enq
follows immediately from \req{S}.  
The equation of motion for the field operator $Q(t)$ is  given by the usual quantum mechanical evolution
\beq  \label{QME}
 i \left[H_t, Q(t) \right]  = \frac{d Q(t)}{d t},
 \enq
with the quantization condition  $[Q(t), P(t)] = [Q(t), \dot Q(t)] = i$. In the Schr\"odinger picture with $t$-independent operators and $t$-dependent state vectors  the evolution is given by
\beq  \label{QE}
H_t \vert \psi(t) \rangle = i \frac {d}{d t} \vert \psi(t)\rangle.
\enq

The absence of dimensional constants in  \req{S}  implies that the action \req{S} and the corresponding equation of motion \req{eom} are invariant under the general group of conformal transformations in the variable $t$, 
\beq \label{ct}
t' = \frac{\alpha t + \beta}{\gamma t + \delta}, \quad  Q'(t') = \frac{Q(t)}{\gamma t + \delta},
\enq
 with $\alpha \delta - \beta \gamma = 1$,
and   there are 
in addition to the Hamiltonian  $H_t$ two more invariants of motion for this field theory, namely the 
dilation operator $D$, and $K$, corresponding to the special conformal transformations in $t$.  
In addition to  $H_t$ \req{Htq}, the operators $D$ and $K$ can be expressed in the field operators $Q(t)$ applying the Noether theorem to the Lagrangian in \req{S}:
\beqa \label{OR}
D &=& t\, H_t -  \quart \left( Q \, \dot Q + \dot Q\, Q\right)  ,\\
K &=& t^2 \, H_t -  \half \, t \left( Q \, \dot Q + \dot Q\, Q\right) +  \half Q^2 \nn ,
\enqa
where we have taken the symmetrized product of the classical expression $Q \dot Q$ since the operators have to be Hermitean.
Using the commutation relations for the field operators $Q(t)$,  one can check that the 
operators  $H_t,  D ~ {\rm and} ~ K$ do indeed fulfill the algebra of the generators of the conformal
group~\cite{deAlfaro:1976je} 
 \beq 
 [H_t,D]= i\,H_t, \quad [H_t ,K]=2\, i \, D, \quad [K,D]=- i\,  K .
\enq

The conformal group in one dimension is locally isomorphic to the group $SO(2, 1)$,  the Lorentz
group in 2+1 dimensions.  In fact, by introducing the combinations
 \beq  \label{a}
 J^{12} =  \half\left(  a \, H_t +\frac{1}{a}  \,K \right) ,\quad J^{01}= \half\left( - a \, H_t +\frac{1}{a}  \,K \right) ,\quad  J^{02} = D,
 \enq
one sees that  the generators $J^{12}$, $J^{01}$ and $J^{02}$ satisfy the algebra of the generators  of the group $SO(2,1)$ 
\beq \label{SO2}
[J^{12},J^{01}]= i\, J^{02}, \quad [J^{12},J^{02}]=- i\, J^{01},
\quad [J^{01},J^{02}]= - i\, J^{12},
 \enq
where $J^{0i},\;  i=1,2$ is the boost in  the space direction $i$ and
$J^{12}$ the rotation in the  (1,2) plane. The rotation operator $J^{12}$ is
compact and has thus a discrete spectrum with normalizable
eigenfunctions. Since the dimensions of $H_t$ and $K$ are different, the  constant $a \ne 0$ has the dimension of $t$.
In fact, the relation between the generators of the conformal group 
and the generators of $SO(2, 1)$ suggests that the scale $a$ may play a fundamental role in physical applications~\cite{deAlfaro:1976je}.
Thus by superposition of different invariants of motion, and thereby introducing a scale, one can hope to come to a confining semiclassical theory, based on an underlying conformal symmetry.

Generally one can construct a  new ``Hamiltonian'' by any superposition of the three constants of motion,
the generators $H_t$, $D$ and $K$,
 \beq \label{G}
 G = u H_t + v D + w K,
 \enq
which is also a constant of motion. The parameters $v$ and $w$, like the constant $a$ in \req{a} are dimensionful. 
Using   \req{QME} and \req{OR} together with the canonical commutation relation for the fields $Q(t)$ we obtain the evolution generated by the operators $D$ and $K$
\beqa
 i \left[D, Q(t) \right]  &=& t \,\frac{d Q(t)}{d t} - \half \, Q(t), \\
  i \left[K, Q(t) \right]  &=&t^2  \frac{d Q(t)}{d t} - t Q(t).
\enqa
Specifically, if one introduces the new variable $\tau$ and the rescaled field $q(\tau)$ defined through~\cite{deAlfaro:1976je}
\beq   \label{qtau}
d\tau= \frac{d t} {u+v\,t + w\,t^2}, \quad q(\tau) = \frac{Q(t)}{(u + v\, t + w \,t^2)^{1/2}},
\enq  
we recover the commutation relation  $[q(\tau), \dot q(\tau)] = i$,  where $\dot q = dq/d\tau$, and
the quantum mechanical evolution for the field operator $q(t)$
\beq   \label{Eq}
 i \left[G, q(\tau) \right]  = \frac{d q(\tau)}{d \tau}.
 \enq
 
 In the Schr\"odinger picture the evolution of the $\tau$-dependent state vectors  is given by  the operator $G$
 \beq    \label{EG}
G \vert \psi(\tau) \rangle = i \frac {d}{d \tau} \vert \psi(\tau)\rangle.
 \enq

If one express the action \req{S} in terms of the transformed fields $q(\tau)$ one finds~\cite{deAlfaro:1976je}
\beq \label{Sq}
S =  \half \int_{\tau_-}^{\tau_+} d \tau \Big( \dot q^2  - \frac{g}{q^2}  - \frac{4 u \omega - v^2}{4} q^2 \Big) + S_{\rm surface}
\enq
where $S_{\rm surface} = \frac{1}{4} \left[\left(v + 2 w t(\tau)\right) q^2(\tau)\right]^{\tau_+}_{\tau_-}$. Expressed in terms of
the original field $Q(t)$ this action \req{Sq},  including the surface terms,  is unchanged.  
However, the  scale factor $4 u w - v^2$ in \req{Sq} breaks the scale invariance of the corresponding Hamiltonian
\beq \label{Htauq}
H_\tau=  \half \Big( \dot q^2 + \frac{g}{q^2} +\frac{4\,uw - v^2}{4} q^2\Big),
\enq
since the surface term is not relevant for obtaining the equations of motion. The Hamiltonian \req{Htauq} is a compact operator  for $4 u w - v^2 > 0$.  It is important to notice that the appearance of the generator of special conformal transformations $K$ in \req{G} is essential for confinement.

We use the Schr\"odinger picture from the representation of $q$ and $p = \dot q$:  $q \to y,  ~ \dot q \to - i {d}/{d y}$,
\beq  \label{Sp}
i \frac{\partial}{\partial \tau} \psi(y,\tau) = H_\tau \Big(y,   - i \frac{d}{d y} \Big) \psi(y,\tau),
\enq
with the corresponding Hamiltonian
\beq  \label{Hx}
H_\tau  = \half \Big(- \frac{d^2}{dy^2} + \frac{g}{y^2} + \frac{4 u \omega - v^2}{4} y^2 \Big),
\enq
in the Schr\"odinger representation~\cite{deAlfaro:1976je}.
For $g \geq -1/4$ and $4\, u w -v^2 > 0$ the 
operator \req{Hx} has a discrete spectrum.

We go now back to the original field operator $Q(t)$ in \req{S}.  From  \req{qtau} one obtains  the relations 
\beq \label{Qq}
q(0) = \frac{Q(0)}{\sqrt{u}},  \quad \quad \dot q(0)= \sqrt{u}\;  \dot Q(0) - \frac{ v}{2 \sqrt{u}}\,Q(0),
\enq
and from \req{Htauq} follows 
\beqa \label{HtauQ}
H_\tau (Q, \dot Q ) \! &= \! & \frac{1}{2} u \left(\dot Q^2 + \frac{g}{Q^2} \right)  - \frac{1}{4} v \left( Q \dot Q + \dot Q Q\right) + \frac{1}{2} w Q^2\\  \nn
            \! &= \! & u H_t + v D + w K,
            \enqa
at $t=0$.  We thus recover the evolution operator \req{G} which describes, like \req{Htauq}, the evolution in the variable $\tau$ \req{EG} but expressed in terms of the original field $Q$.

With the realization of the operator $Q(0)$ in the state space  with wave functions $\psi(x, \tau)$ and the substitution   $Q(0) \to x$ and $\dot Q(0) \to - i \frac{d}{d x}$ we obtain
\beq  \label{Sp}
i \frac{\partial}{\partial \tau} \psi(x,\tau) = H_\tau \Big(x,   - i \frac{d}{d x} \Big) \psi(x,\tau),
\enq
and from \req{HtauQ}  the Hamiltonian
\beq \label{Htaux}
H_\tau= \frac{1}{2} u \left(- \frac{d^2}{d x^2} + \frac{g}{x^2}\right)  + \frac{i}{4} v  \left(x \, \frac{d}{d x}+ \frac{d}{d x} \, x  \right) +\frac{1}{2}  w x^2.
\enq

The field $q(\tau)$ was only introduced as an intermediate step in order  to recover the evolution equation \req{Eq} and the Hamiltonian \req{Htauq}
from the action \req{S}  with variational methods.  This shows that the essential point is indeed the change from $t$ to $\tau$ as the evolution parameter.

\section{Connection to Light-Front Holographic QCD}

We are now in a position to apply the group theoretical results from the conformal algebra to the front-form ultra-relativistic bound-state wave equation. Comparing the Hamiltonian \req{Htaux}  with the light-front wave equation \req{LFWE} and identifying the variable $x$ with the light-front invariant variable $\zeta$,  we have to choose $u=2, \; v=0$ and relate the dimensionless constant $g$ to the LF orbital angular momentum, $g=L^2-1/4$,  in order to reproduce the light-front kinematics. Furthermore  $w = 2 \lambda^2$ fixes the confining light-front  potential to a quadratic $\la^2 \, \zeta^2$ dependence.  From  \req{Pm} and \req{Sp} we find $\tau = x^+/P^+$ in the 
LF frame $\vec P_\perp = 0$.  The dimension of $\tau$ is that of an inverse mass squared, characteristic of the fully relativistic treatment.

The relation with the algebra of the group $SO(2,1)$ becomes particularly compelling when  the Hamiltonian $H_\tau$ \req{Htaux} is mapped to the light-front. In this case one has $u=2$, $v=0$, and $w =  2 \la^2$ and the LF Hamiltonian can be expressed in terms of the generators  $H_t$ and $K$
 \beq \label{HtauLF}
 H_\tau^{LF}= 2 \left( H_t + \lambda^2 K\right),
 \enq
 From the relations \req{a}  follows the connection of the free Hamiltonian $H_t$, \req{Htq} with the group generators of $SO(2,1)$
\beq \label{free}
J^{12} -J^{01}= a H_t.
\enq
The Hamiltonian  \req{HtauLF}  can be expressed as a generalization of \req{free} by replacing $J^{12} - J^{01}$ by $ J^{12} - \chi J^{01}$. This generalization yields indeed
\beq \label{conf}
 J^{12} - \chi  J^{01}
= \frac{1}{4} {a}(1+\chi) H_\tau^{LF},
\enq
with
\beq \label{aw}
\la^2=\frac{1}{a^2} \, \frac{1-\chi}{1+\chi},
\enq
in \req{HtauLF}.  For $\chi=1$ we recover the free case \req{free}, whereas for $ -1 <\chi < 1$ we obtain a confining LF potential.  For $\chi$ outside this region, the Hamiltonian is not bounded from below.

This consideration based on the isomorphism of  the conformal group in one dimension with the group $SO(2,1)$ makes  the appearance of a dimensionful constant in the Hamiltonian \req{HtauLF} derived from a conformally invariant action less astonishing. In fact,  as mentioned below
Eqs. \req{a} and \req{SO2}, one has to introduce the dimensionful constant $a \ne 0$ in order to relate the generators of the conformal group with those of the group $SO(2,1)$.  This constant $a$  sets the scale for the confinement strength $\la^2$ but does not determine its magnitude, as can be seen from \req{aw}. This value depends on  $a$ as well as on the relative weight of the two generators $J^{12}$ and $J^{01}$ in the construction of the Hamiltonian \req{conf}.

Comparing the Hamiltonian \req{HtauLF}   with the  potential \req{UU} derived from the modified AdS geometry \req{U}, we see that we are indeed fixed to a 
confining term proportional to $\zeta^2$, corresponding to a quadratic dilaton profile $\varphi = \la z^2$, with $\la = \sqrt{w/2}$. We obtain from \req{U}  the LF  potential \req{UU}.

\section{Conclusions}

The main result of this Letter is the unique determination of the confining potential from the underlying conformal symmetry of the classical QCD Lagrangian, incorporated in an effective one-dimensional light-front effective theory for QCD. In fact, using  the triple complementary connections of  AdS space, its LF holographic dual and the relation to the algebra of the conformal group in one dimension we find a dilaton profile $z^s$ with the unique power $s = 2$. This  has important  phenomenological consequences.   Indeed,  for $s=2$ the mass of the $J=L=n=0$ lowest bound state, which we identify with the pion, is automatically zero in the chiral limit, and the separate dependence on $J$ and $L$ leads to a  mass ratio of the $\rho$ and the $a_1$ mesons which coincides with the result of the Weinberg sum rules. One predicts linear  Regge trajectories~\cite{Karch:2006pv}  with the same slope in the relative orbital angular momentum $L$ and the radial quantum number $n$.  In the phenomenological application the constant term $2 \la (J-1)$ in the potential, which is not fixed by the group theoretical arguments, is crucial. This term is fixed by the holographic duality to light-front  quantized QCD~\cite{deTeramond:2013it}.
It can be shown that only the power $s=2$ in the dilaton term leads to linear trajectories and to a massless pion but without apparent breaking of chiral symmetry.
In addition, since quantum fluctuations have no infrared divergences because of color confinement, the gap parameter $\Lambda^2_{QCD}$ appearing in the evolution of the QCD running coupling will be determined in the present framework by the mass scale $\lambda$.

The dAFF mechanism for introducing a scale makes use of the algebraic structure of one-dimensional conformal field theory.  A new Hamiltonian with a mass scale $\sqrt{\la}$ is constructed from the generators of the conformal group and its  form is  therefore fixed uniquely: it is, like the original Hamiltonian with unbroken dilatation symmetry, a constant of motion~\cite{deAlfaro:1976je}.  The essential point of this procedure is the introduction of a new evolution parameter $\tau$.  The local isomorphism between the conformal  group in one dimension and the  group $SO(2,1)$ is fundamental for introducing the  scale for confinement in the light-front Hamiltonian and fixing the dilaton profile in  AdS.
The dAFF mechanism is reminiscent of spontaneous symmetry breaking, however, this is not the case  since there are no degenerate vacua~\cite{Fubini:1984hf} and thus a massless scalar $0^{++}$ state is not required. The dAFF mechanism is also different from  usual explicit breaking by just adding a term to the Lagrangian~\cite{Kharzeev:2008br}.

A specific value for $\lambda$ is not determined by QCD alone. The scale only becomes fixed when we make a measurement such as the pion decay constant or the $\rho$ mass. The scale $\Lambda^2_{\rm QCD}$ appearing in the QCD running coupling $\alpha_s(Q^2)$ can be determined from the fundamental QCD confinement scale $\lambda$ by matching the perturbative and nonperturbative regimes.  The specific value of  $\Lambda^2_{\rm QCD}$ depends on the choice of the renormalization scheme for $\alpha_s(Q^2)$, such as the $\overline{MS}$ scheme.

In their discussion of the evolution operator $H_\tau$ 
dAFF mention a critical point, namely that ``the time evolution is quite different from a stationary one''.  By this statement they refer to the fact that the variable $\tau$ is related to the variable $t$  by 
\beq
 \tau= \frac{1}{\sqrt{ 2w}}\, \arctan \left(t \sqrt{\frac{w}{2}}\right),
\enq
{\it i.e.}, $\tau$ has only a finite range.  In our approach $\tau = x^+/P^+$ can be interpreted as the LF time difference of the confined $q$ and $\bar q$ in the hadron, a quantity which is naturally of finite range and in principle could be measured in double-parton scattering processes.

It is  interesting to notice that since the algebra of the conformal group in one dimension with generators $H_t$, $K$ and $D$ is locally isomorphic to the group $SO(2,1)$, a correspondence can be established  between the $SO(2,1)$ group of conformal quantum mechanics and the AdS$_2$ space with isometry group $SO(2,1)$~\cite{Chamon:2011xk}. 
Thus, we cannot expect that the dilaton field in the modified AdS$_5$ action could be scaled away by some redefinition of the variables  to recover pure AdS$_5$, since the fundamental conformal  invariance with $SO(2,1)$  generators operates  on a dual subspace, namely AdS$_2$.

Following the remarkable work of de Alfaro, Fubini and Furlan in Ref.~\cite{deAlfaro:1976je}, we have discussed in this Letter an effective theory which encodes the fundamental conformal symmetry of the QCD Lagrangian in the limit of massless quarks. It is an explicit model in which the confinement length scale appears in the light-front Hamiltonian   from the breaking of dilatation invariance, without affecting  the conformal invariance of the action. In the context of  the dual holographic model  it shows that the form of the dilaton profile is unique, which leads by the mapping to the  light-front Hamiltonian to a unique confining potential.

Our approach is very different than the standard interpretation of broken chiral symmetry. The pion bound-state solution corresponds to a composite $q \bar q$ state, not an elementary field.    The specific vanishing of the pion mass  in the chiral limit follows from the precise cancellation of light-front kinetic energy and potential energy without apparent chiral symmetry breaking~\cite{Brodsky:2009zd,  Brodsky:2010xf}.  Thus the model described here provides a new insight in the zero mass pion  chiral limit of QCD, which is not necessarily in conflict with the standard viewpoint of chiral symmetry.  As in the original work of Polchinski and Strassler~\cite{Polchinski:2001tt}, the pion form factor falls off as $1/ Q^2$, consistent with  its leading twist-two  interpolating operator  at short distances. The shape of the pion form factor agrees with experiment.  The prediction for the ratio of the masses of the $a_2$ and $\rho$ mesons agrees with Weinberg's sum rule. Furthermore, the systematics of the radial and orbital excitations predicted by the model agree with experiment, a prediction which has not been possible within the context of current algebra alone.  The model described in this Letter provides a compelling effective frame-independent effective theory of hadrons since it predicts essential features of the hadron spectrum, including the masslessness of the pion, and provides an interesting alternative approach to the conventional descriptions of nonperturbative QCD.

The main result of this Letter is that group theoretical arguments based on the underlying conformality of the classical Lagrangian of QCD fix the quadratic form  of the effective dilaton profile and thus the corresponding form of the confinement potential of the light-front QCD Hamiltonian. Though this is particularly important for light-front holographic QCD, it is also relevant to other bottom-up holographic QCD models, as for instance the model discussed in~\cite{Karch:2006pv}.

\section*{Acknowledgements}

\noindent We thank Diego Hofman for pointing out the work of de Alfaro, Fubini, and Furlan  to us and to William Bardeen, Alonso Ballon-Bayona, James Bjorken, Joseph Day, Paul Hoyer, Roman Jackiw, Shamit Kachru,  Ami Katz, Matin Mojaza, Dieter M\"uller, Volkhard M\"uller,  Michael Peskin, Eliezer Rabinovici  and Robert Shrock   for  valuable discussions.
This work  was supported by the Department of Energy contract DE--AC02--76SF00515.

\end{document}